# SYNTHESIS, CHARACTERIZATON, AND APPLICATION OF CARBON NANOTUBES

*Kuen-Song Lin[*], Ruey-Bing Li, Frank Lee, Li-Cheng Lou, and Su-Wei Chiu*

Department of Chemical Engineering & Materials Science, Yuan-Ze University, Chung-Li, Taiwan 320, R.O.C.

**ABSTRACT**

Carbon nanotubes (CNTs) have attracted increasing attention because of their unique structural, mechanical, and electronic properties. Surface chemistry modifications are also useful and critical to manipulate the adsorptive properties of CNTs and develop their potential of environmental application. Therefore, the main objectives of this study were to investigate the optimal synthesis methods or characteristics identification of multiwall CNTs (MWCNTs) and the environmental application of lead ions adsorptive removal by CNTs. Experimentally, the MWCNTs of well-graphited walls were obtained with reductive potassium metals of catalytic hexachlorobenzene-thermal routes in the presence of Co/Ni catalyst precursors at 503-623 K for 8-12 hr. The diameters of MWCNTs ranged of 30-100 nm by TEM microphotos. In addition, fine structures or surface chemical modification of MWCNTs or Co/Ni precursors were further identified by EDS, XRPD, X-ray absorption near edge structural (XANES) or extended X-ray absorption fine structural (EXAFS) spectroscopy.

## 1. INTRODUCTION

Carbon nanotubes (CNTs) synthesized in the arc-discharge apparatus was first discovered by Iijima in 1991 as nested structures of concentric shells, which stimulated a worldwide research effort to improving their synthesis, structure determination, and physical properties calculation and identification [1]. The unique carbon nanotubes are predicted to possess novel mechanical and electrical properties since their regular, periodic structure, and quantum size [2-4]. These properties suggest enormous potential applications for CNTs include such as novel nanoscale electronic devices, tips for scanning probe microscopy, hydrogen storage media, reinforcing materials for carbon matrix composites, and catalyst supports [3-6]. Realization of these commercial scale uses will require scaleable, economical synthesis methods. In the recent years, methods for synthesizing metal-catalyzed CNTs has been pursued by a variety of methods including laser or arc evaporation of a metal praphite composite target, hydrocarbon pyrolysis or CO disproportionation on a metal catalyst in high temperatures [7-11]. The laser or arc evaporation provides little control over process condition, thus producing many undesired by-products, a low CNTs yield, and posing physical separation challenges [12-16]. In contrast, high temperature furnaces offer a high degree of control over the hydrocarbon identity, temperature, and reaction time, all of which can be optimized for product purity [17-21]. However, these methods are energy intensive and produce relatively low quantities of CNTs. In addition, neither method is readily scaleable for bulk commercial production.

On the contrary, a novel and innovatory catalytic-assembly benzene-solvothermal route to MWCNTs by using a reduction of hexachlorobenzene on metallic K in the presence of Co/Ni catalyzer in a autoclave has been used. Comparatively the reaction temperature of this CNTs synthesis system is much lower than others. The valency and fine structure of Co and Ni atom used as the catalyzer precursor of CNTs formation have also not been well studied. The concurrence of nucleation of the Co/Ni metal nanoparticles and CNTs growth within a reactive hydrocarbon atmosphere creates a complex transient chemical environment, making it difficult to study the inception and growth processes in CNTs synthesis.

By using XANES/EXAFS spectroscopy, the valency and fine structure of these complex metals in the synthesis of CNTs has been investigated. The XANES/EXAFS spectra offer the basic knowledge of understanding the oxidation states and fine structures of Co and Ni atom in the complex catalytic precursor to a further study on the distribution/dispersion and interaction mechanism of these metals. XANES/EXAFS spectroscopy is also an excellent technique for characterizing the valency and local structure of Co/Ni metals in catalyzer precursor with short-range order [22-25]. Thus, the main objective of the present study was to investigate the fine structures, oxidation states, fine structures or surface chemical modification of MWCNTs and Co/Ni atoms dispersed in the catalyzer precursor of MWCNTs synthesis by TEM, XRPD, EDS or XANES/EXAFS spectroscopy. The chemical-modified with concentrated $H_2SO_4/HNO_3$ solution of MWCNTs used for the adsorptive enhancement of Pb(II) removal was also studied.

## 2. EXPERIMENTAL

The MWCNTs synthesis was conducted in a batch system. 100 mL of benzene was placed into a 250 ml PTFE-lined stainless steel 316 autoclave, and then 15 grams hexachlorobenzene and 20 grams metallic potassium were added. After that, 15 mL of benzene was added until approximately 70-80% of the autoclave's capacity was filled. Finally, 700 mg of catalyzer precursor were also added. The catalyzer precursor was prepared by dissolving one gram of $CoCl_2 \cdot 6H_2O$ and $NiCl_2 \cdot 6H_2O$ mixture (Co/Ni = 1 in mole ratio) in 100 mL of absolute ethanol, followed by drying in an 353 K and calcining at 423 K in a flowing air. All the chemicals used are high-purity (ACS grade). The sealed autoclave was heated at 503-623 K for 8-12 hr, and then cooled to room temperature in the furnace naturally. The obtained samples were repeatedly and sequentially washed with absolute ethanol, dilute acid, de-ionized double distilled water to remove the residual impurities, such as chlorides and remaining catalyzer, and finally vacuum-dried at 343-353 K for 6-10 hr.

In order to more thoroughly examine the adsorptive efficiencies of 50-100 mg/L $Pb^{2+}$ on K-MWCNTs, the batch adsorption experiments at 298 K were conducted. The average Co/Ni metal contents in the catalyzer precursor were evaluated by atomic absorption spectroscopy (AAS, GBC model 908) and induced couple plasma/mass spectroscopy (ICP/MS, ELAN model 5000). Each calibration curve was generated with its corresponding standard metal solution at ten different concentrations bracketing the expected Co/Ni metal concentrations in the catalyzers. Concentrations of Co/Ni metals in the catalyzer precursors were then calculated. The morphologies of the samples were observed with Model Zeiss 10C transmission electron microscope (TEM) performed at 200 kV, and the microstructure of MWCNTs were measured with JEOL 2010 high-resolution transmission electron microscope (HRTEM) conducted also at 200 kV. Samples for the electron microscope were prepared by 1 h ultrasonic dispersion of 0.2 g of product with 50 mL of absolute ethanol in a 100 mL conical flask. Then a drop of the solution was placed on a copper microgrid or carbon film and dried in air before performance. Structures of MWCNTs or Co/Ni precursors samples were measured by X-ray powder diffraction (XRPD) scanned from 10 to 90°($2\theta$) with a scan rate of 4°($2\theta$) $min^{-1}$ with monochromatic $CuK_\alpha$ radiation (MAC Sience, MXP18).

The XANES/EXAFS spectra were collected at the Wiggler beamline 17C1 at the National Synchrotron Radiation Research Center (NSRRC) of Taiwan. The electron storage ring was operated with an energy of 1.5 GeV and a current of 100-200 mA. A Si(111) DCM was used for providing highly monochromatized photon beams with energies of 1 to 15 keV and resolving power (E/$\Delta$E) of up to 7000. Data were collected in fluorescence or transmission mode with a Lytle ionization detector [25] for Co (7709 eV) and Ni (8333 eV) K edge experiments at room temperature. The photon energy was calibrated by characteristic preedge peaks in the absorption spectra of Co and Ni standards. The raw absorption data in the region of 50 to 200 eV below the edge position were fit to a straight line using the least-square algorithms. The fitted pre-edge background curves were extrapolated throughout all data range, and subtracted and normalized to minimize the effect of the MWCNT sample thickness. The near-edge structure in an absorption spectrum covers the range between the threshold and the point at which the EXAFS begins [22-24]. The XANES extend to an energy of the order of 50 eV above the edge. The $k^2$-weighted and EXAFS spectra were Fourier transformed to $R$ space over the range between 2.5 and 12.5 $Å^{-1}$. The EXAFS data were analyzed by using the UWXAFS 3.0 program and FEFF 8.0 codes [22-25].

## 3. RESULTS AND DISCUSSION

The TEM microphotographs were used to investigate the morphologies, crystallinity, and microstructure of MWCNTs synthesized by using Co/Ni catalyzer precursors with metallic K. Morphological observations of MWCNTs synthesized by using Co/Ni catalyzer precursors were shown in Figure 1. Figure 1(a) showed the MWCNTs have a bamboo-like structure with an average length of 1500 nm. The inner diameter of MWCNTs is 20 nm and outer diameter 45 nm in average. The inner tube is subdivided by single- or multi-graphite layer, which may be formed by the surface diffusion of carbon cluster on the larger Co/Ni catalytic nanoparticles across the wall of carbon nanotubes. The MWCNTs having irregular-shaped were showed in Figure 1(b). The MWCNTs without amorphous carbon coating typically have open ends when were acid- or ultrasonic-purified previously.

Figure 2 represented the metallic K was found in MWCNTs synthesized by using Co/Ni catalyzer precursors at T = 503-623 K for 12 h identified with energy dispersive spectrometer (EDS). The postulated equation and reaction mechanism of MWCNTs synthesis may be described as following:

n $C_6Cl_6$ + n K $\rightarrow$ 6n KCl + poly aromatic hydrocarbons (+ benzene at 503-623 K) $\rightarrow$ MWCNTs (+ Co/Ni for 8-12 hr)

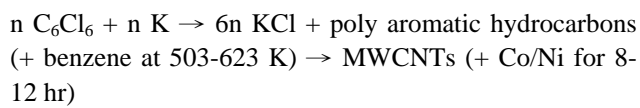

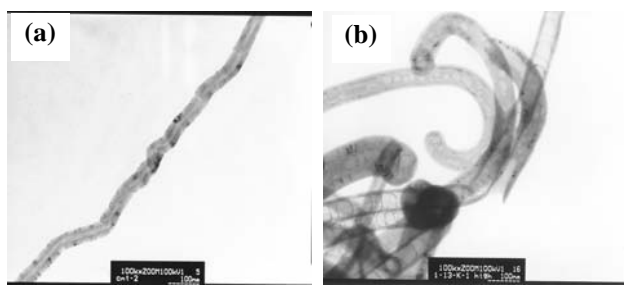

Figure 1 TEM images of (a) bamboo-like and (b) irregular-shaped of MWCNTs synthesized by using Co/Ni catalyzer precursors with metallic K at T = 503-623 K for 12 hr.

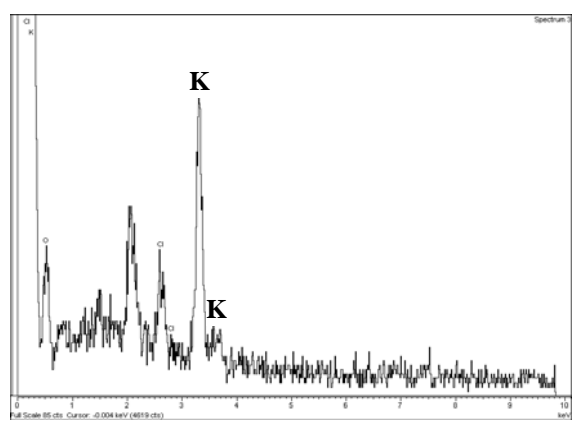

Figure 2 EDS measurement of MWCNTs synthesized by using Co/Ni catalyzer precursors with metallic K at T = 503-623 K for 12 hr.

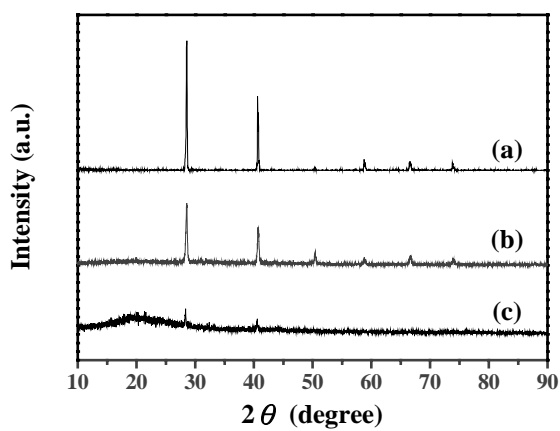

Figure 3 XRPD patterns of (a) KCl standard, (b) unwashed MWCNTs with KCl crystals, (c) washed MWCNTs with KCl contaminants synthesized by Co/Ni catalyzer precursors at T = 503-623 K for 12 h.

The XRPD pattern shown in Figure 3 indicated that the abstraction of Cl from hexachlorobenzene and formation of the KCl precipitate may be involved in the early stage of the synthesis process of MWCNTs. Since KCl powders can be easily dissolved in water, the high removal efficiency of KCl from MWCNTs samples by washing with water was found (shown in Figures 3 (b) and (c)). KCl was the main species and over 95% of $K^+$ cations was formed as KCl in the synthesis process of MWCNTs by using Co/Ni catalyzer precursors at at T = 503-623 K for 8-12 hr.

Since the valency and fine structure of Co and Ni atom used as the catalyzer precursor of MWCNTs formation have not been well studied, therefore the catalytic redox mechanisms effected by the Co/Ni catalyzer precursor in the growth of MWCNTs may be determined by using XANES and EXAFS spectra. Figure 4 showed that the XANES or EXAFS data may indicate the chlorides of Co and Ni will be reduced to Co/Ni catalyzer particles by metallic potassium, at the same time, hexachlorobenzene may be also reduced by potassium through deleting chlorides to form carbon clusters and KCl. By using XANES spectra of Co/Ni precursor residues shown in Figure 4 (a), the Co/Ni catalyst precursors of the MWCNTs syntheses were all metallic Co or Ni species. In addition, all the adsorption curves of Co/Ni precursor residues were compared and confirmed with the ones of metallic Co or Ni powder standards.

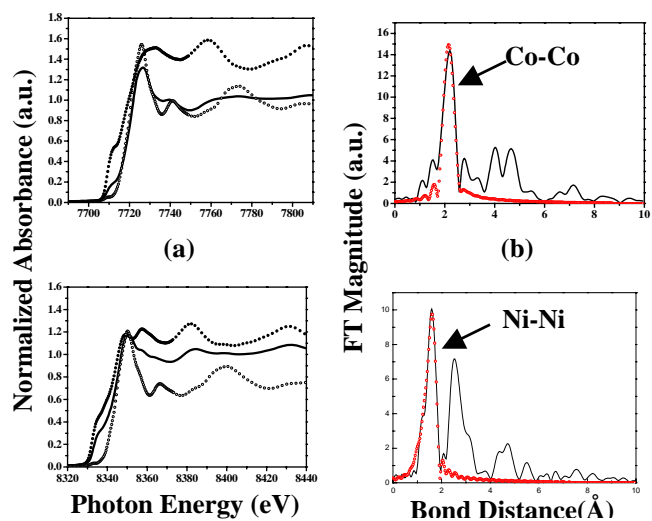

Figure 4 (a) NANES of Co/Ni precursor residue samples (solid lines) compared with Co(0,II) or Ni(0,II) powder standards and (b) Fourier transform (FT) of the metallic Co and Ni K-edge EXAFS of the carbon nanotubes synthesized by using Co/Ni catalyzer precursors at T = 603 K for 8-12 hr. The best fitting of the EXAFS spectrum is expressed by the dotted lines.

The EXAFS data of Co/Ni precursor residues revealed that the nanophase metallic Co or Ni particles had a central Co (or Ni) atom of a coordination number of 2.92 ± 0.05 (or 5.96 ± 0.05) primarily Co-Co with a bond distance of 2.49 ± 0.05 Å (or 2.48 ± 0.05 Å). These results may offer a further explanation of how the yield and structure of MWCNTs formed by catalytic-assembly benzene-solvothermal route depend on the species of the Co/Ni metal catalysts. However, this result combining with TEM images (Figures 1(a) and (b)) might indicate that some MWCNTs having bending structure, and at the close end of the tubes, the catalytic Co/Ni metal particles exist clearly, which may suggest that the Co/Ni metal particles are responsible for the nucleation of the MWCNTs. The concurrence of nucleation of the Co/Ni metal nanoparticles and MWCNTs growth within a reactive hydrocarbon atmosphere (e.g. benzene) creates a complex transient chemical environment, making it difficult to study the inception and growth processes in MWCNTs synthesis.

Since the unique adsorptive properties of carbon nanotubes, it suggests enormous potential applications for MWCNTs on environmental technologies. Adsorptive removal of toxic metals such as Pb(II) cations by surface-modified MWCNTs is an appealing pollution control and metallic resource recovery alternative. Since the sake of chemical decoration and modification of K-MWCNTs, the first cap opening of MWCNTs may be finished and therefore the certain active functional groups are observed. Laboratory-scale experiments were carried out to provide information concerning the feasibility for Pb(II) adsorption system of MWCNTs, and determined if further developments would be warranted.

In order to more thoroughly examine the adsorptive efficiencies of 50-100 mg/L Pb(II), different oxidants of chemical modifications of MWCNTs surfaces for adsorption enhancement included $HNO_3$, $KMnO_4$ or $Na_2S$ processes were performed. Experimentally, the Pb(II) adsorptive efficiencies of K-MWCNTs were $KMnO_4$ > $HNO_3$ > $Na_2S$ modification processes in series. Interestingly, these methods are all effective in opening the caps of MWCNTs but are very different in oxidation ability. The surface roughness and abundance of effective functional groups such as OH, COOH or NO on the MWCNTs after surface modification also vary. Since using $KMnO_4$ in alkali solution as a moderate oxidant, the caps of MWCNTs may be not effectively opened. Similarly, dilute $HNO_3$ solution may purify the MWCNTs and etch the nanotubes from the terminal and the defects on the nodes or sidewalls. However, the relatively mild oxidant of $HNO_3$ solution may not able to generate abundant NO, OH or COOH functional groups eventually.

The experimental results of higher Pb(II) adsorptive removal efficiency of MWCNTs indicated that the $KMnO_4$ may provide abundant and more effective –OH or COOH functional groups on MWCNTs. Moreover, the MWCNTs possessed COOH functional groups on the surface of defect structures during chemical oxidation of $HNO_3$ and the –NO functional groups were observed on the K-MWCNTs surface of defect structures.

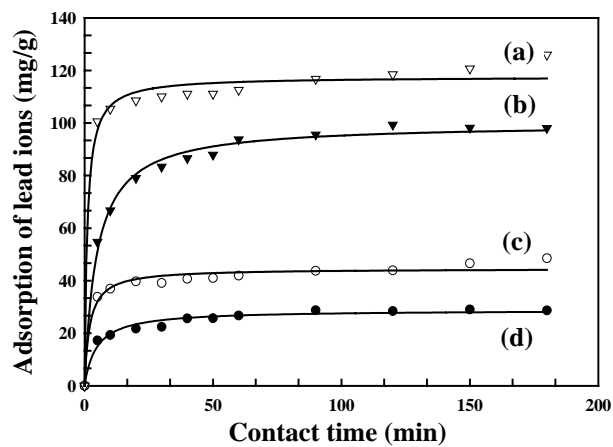

Figure 5 The adsorption curves of 100 mg/L Pb(II) on surface-modified MWCNTs with (a) $KMnO_4$, (b) $HNO_3$, (c) $Na_2S$, and (d) non-modification procedures.

## 4. CONCLUSIONS

The potential synthesis route of MWCNTs with reductive metallic K of catalytic hexachlorobenzene-thermal routes in the presence of Co/Ni catalyst precursors at 503-623 K for 12 hr was investigated. The characteristics identification of as-synthesized MWCNTs and environmental application of lead cations adsorptive removal by MWCNTs were also studied. Experimentally, TEM microphotos showed the MWCNTs have a bamboo-like structure with an average length of 1500 nm. The inner diameter of CNTs is 20 nm and outer diameter 45 nm in average. The XRPD patterns indicated that the abstraction of Cl from hexachlorobenzene and formation of the KCl precipitate may be involved in the early stage of the synthesis process of MWCNTs. From the XANES spectra, the Co/Ni catalyst precursors of the MWCNTs syntheses were all metallic Co or Ni species. Furthermore, the EXAFS spectra of Co/Ni precursor residues revealed that the metallic Co or Ni nanoparticles had a central Co (or Ni) atom of a coordination number of 2.92 ± 0.05 (or 5.96 ± 0.05) primarily Co-Co with a bond distance of 2.49 ± 0.05 Å (or 2.48 ± 0.05 Å). Finally, in the view of environmental application, the Pb(II) adsorptive efficiencies of MWCNTs were $KMnO_4$ > $HNO_3$ > $Na_2S$ modification processes in series.